\documentclass[10pt,a4paper]{article}

\usepackage[T1]{fontenc}
\usepackage{hyperref}
\usepackage{graphicx,subfigure}
\usepackage{amsmath,amssymb,upgreek}
\usepackage{psfrag}
\usepackage[font=footnotesize,labelfont=bf]{caption}

\usepackage{enumerate}

\usepackage[left=2cm,top=2.35cm,right=2cm,bottom=2.35cm]{geometry}

\begin{document}
\begin{center}
\Large{\textbf{Probing kinematics and fate of the Universe with linearly time-varying deceleration parameter}} \\[0.5cm]
 
\large{{\"{O}zg\"{u}r Akarsu$^{\rm a}$,  Tekin Dereli$^{\rm a}$, Suresh Kumar$^{\rm b}$,   Lixin Xu$^{\rm c}$}}
\\[0.5cm]

\small{
\textit{$^{\rm a}$ Department of Physics, Ko\c{c} University, 34450 Sar{\i}yer, {\.I}stanbul, Turkey.}}

\vspace{.2cm}

\small{
\textit{$^{\rm b}$ Department of Mathematics, BITS Pilani, Pilani Campus, Rajasthan-333031, India.}}

\vspace{.2cm}

\small{
\textit{$^{\rm c}$ Institute of Theoretical Physics, Dalian University of Technology, Dalian, 116024, P. R. China.}}

\end{center}
\textbf{E-Mail:} oakarsu@ku.edu.tr, tdereli@ku.edu.tr, suresh.kumar@pilani.bits-pilani.ac.in, lxxu@dlut.edu.cn

\vspace{.8cm}

\noindent {\textbf{Abstract.} 
{\small The parametrizations $q=q_0+q_1 z$ and  $q=q_0+q_1 (1-a/a_0)$ (Chevallier-Polarski-Linder parametrization) of deceleration parameter, which are linear in cosmic redshift $z$  and scale factor $a$, have been  frequently utilized in the literature to study kinematics of Universe. In this paper, we follow a strategy that leads to these two well known parametrizations of deceleration parameter as well as an additional new parametrization $q=q_0+q_1(1-t/t_0)$, which is linear in cosmic time $t$. We study the features of this linearly time-varying deceleration parameter in contrast with the other two linear parametrizations. We investigate in detail the kinematics of the Universe by confronting the three models with the latest observational data. We further study the dynamics of the Universe by considering the linearly time-varying deceleration parameter model in comparison with the standard $\Lambda$CDM model. We also discuss future of the Universe in the context of the models under consideration.
}
\vspace{0.3cm}

\section{Introduction}
\label{sec:Intro}

It was thought that the expansion of the current Universe could be well described by solving Einstein's field equations in the presence of pressure-less matter (dust) within the framework of spatially flat Robertson-Walker spacetime, which gives a decelerating expansion with a constant deceleration parameter (DP) equal to $\frac{1}{2}$. The experimental efforts to confirm this, however, led to the discovery that the current Universe is in fact accelerating \cite{Riess98,Perlmutter99}. It is today very well established that the Universe is not only expanding but also has evolved from decelerating expansion to accelerating expansion, and is in the accelerating expansion phase during the last $\sim 6$ billion years. However, we still have no satisfactory explanation for this fact that occurs at energy scales $\sim 10^{-4}\,{\rm eV}$, where we supposedly know physics very well. This accelerated expansion of the Universe, essentially, requires either the presence of an energy source in the context of Einstein's general theory of relativity whose energy density is constant/varies very slowly as the Universe expands and that permeates all over the space uniformly (dubbed as dark energy) \cite{Copeland06,Bamba12} or a modification of the Einstein's general theory of relativity for describing gravitation at cosmological scales \cite{Capozziello11extended,Nojiri11,Clifton12}.

The most successful cosmological model, we know so far, is the $\Lambda$CDM model that is obtained simply by the inclusion of a positive cosmological constant $\Lambda$ (which is mathematically equivalent to the conventional vacuum energy density that is described with an equation of state (EoS) parameter equal to $-1$) into the Friedmann equations for spatially flat Universe containing dust \cite{GronHervik}. However, it suffers from two serious conceptual/theoretical problems known as the fine tuning and coincidence problems \cite{Zeldovich,Weinberg89,Sahni00}, which led to an immense search for possible alternatives to $\Lambda$CDM model starting right after the first discovery of the current acceleration of the Universe. On the other hand, latest cosmic microwave background (CMB) experiment, the \textit{Planck} experiment whose major goal is to test $\Lambda$CDM model to high precision and identify areas of tension, shows that the \textit{Planck} data are remarkably consistent with the predictions of the base $\Lambda$CDM model \cite{PlanckCosmological}. However, the \textit{Planck} group also announces some issues; for instance, the presence of a mismatch with the temperature spectrum and some other anomalies at low multipoles $l\lesssim 30$; the strikingly low value of the Hubble constant $H_{0}=67\pm 1.4\,{\rm km}\,s^{-1}\,{\rm Mpc}^{-1}$ (which is in tension at about 2.5$\sigma$ level with the direct measurements of Hubble constant $H_{0}=73\pm 2.4\,{\rm km}\,s^{-1}\,{\rm Mpc}^{-1}$ \cite{Riess11} and cannot be resolved by varying the parameters of the $\Lambda$CDM model and is also in mild tension at about 2$\sigma$ level with the Supernova Legacy Survey (SNLS) compilation while it is compatible with the Union2.1 compilation whose scatter, however, is significantly larger than that of the SNLS compilation). \textit{Planck} group concludes that the tension between CMB-based estimates and the astrophysical measurements of $H_{0}$ is intriguing and merits further discussion. Regarding the dark energy, it is found that while the CMB data alone is compatible with cosmological constant assumption in the base $\Lambda$CDM model, the addition of astrophysical data into the analysis usually draws the EoS of the dark energy into the phantom domain ($w<-1$), which motivates to think of possible time dependence of $w$. Accordingly, the \textit{Planck} group considers the simple linear relation $w(a)=w_{0}+w_{a}(1-a)$, and concludes that the dynamical dark energy is favored at about $2\sigma$ level when the direct measurement of $H_{0}$, or the SNLS SNe sample, together with \textit{Planck} data are considered. Such \textit{Planck} experiment results motivate the inquiries for possible alternatives that do not deviate from the $\Lambda$CDM model much but may fit the observational data better and relieve the tensions as mentioned above. 

In the analyses of cosmological data it is usually assumed that general relativity is valid at cosmological scales and the physical ingredient of the current Universe is made of a dark energy source and dust or that a modified gravity is valid at cosmological scales so that an additional dark energy source will not be necessary. Such analyses are often referred to as \textit{dynamical studies}. One of the advantages of such studies is that the cosmological model that will be confronted with the observations can usually be constructed starting from a well motivated theoretical background. However, the different cosmological models usually give different results for the kinematics of the Universe even in the range covered by the considered observational data, and can make us predict completely different histories for the Universe out of the range covered by the considered observational data. Moreover, we are not sure yet that whether the general relativity needs to be modified or the presence of an unknown dark energy source is inevitable. Hence, it makes sense to try to obtain information on the kinematics of the Universe with no assumption for a particular metric theory of gravity or for the matter-energy content of the observed Universe. It should be admitted that such a route has less chance to shed light on the question of validity of general relativity at cosmological scales and the nature of the dark energy as well. However, this route gives us opportunity to investigate the history of the Universe in a less model dependent way/more directly, and to consider a larger class of expansion histories. Besides, this route is in fact not totally divorced from the dynamical studies. The considered parametrization for the kinematics of the Universe can often be converted into a dynamical set of equations by assuming a particular metric theory of gravity (for instance general relativity), and then the properties of the physical ingredient of the Universe can be investigated. A crucial point, when this route is taken up, is to develop a reasonable strategy for parametrizing the kinematics of the Universe. Various strategies can be followed depending on some practical, mathematical, phenomenological, theoretical and other aspects \cite{Turner,Rapetti07,Ishida08,Clarkson09,Xu09,Lu11,Barboza12,Li12fate,Farooq13}.

Three important parameters in cosmology are the Hubble parameter $H$, dimensionless DP $q$ and the dimensionless jerk parameter $j$ that are defined as follows:
\begin{equation}
H=\frac{\dot{a}}{a},\quad q=-\frac{\ddot{a}a}{{\dot{a}}^{2}}\quad\textnormal{and}\quad j=\frac{\dot{\ddot{a}}a^2}{\dot{a}^3},
\end{equation}
where $a$ is the scale factor and an over dot denotes derivative with respect to cosmic time $t$. DP among these is a key quantity in describing the evolution of the Universe. It not only characterizes decelerating ($q>0$)/accelerating ($q<0$) expansion of the Universe but also gives the EoS parameter  $w=p/\rho$ (where $\rho$ and $p$ are the energy density and pressure respectively) of the effective cosmic fluid through the relation $q=\frac{1}{2}+\frac{3}{2}w$ in general relativity within the framework of spatially flat RW spacetime.  Accordingly, before the first indications of the current acceleration of the Universe, in 1998, the DP of the Universe was expected to be $q\cong\frac{1}{2}$ since it was believed that the physical ingredient of the current Universe could be described very well with a dust with a negligible EoS parameter $w\cong0$. Also, constant DP leads to power-law or exponential-law cosmologies, which are not suitable for describing the dynamical evolution of the Universe \cite{Kumar12}. In the standard $\Lambda$CDM model that fits the data successfully, the DP is variable and evolves from $\frac{1}{2}$ to $-1$ while the jerk parameter is simply a constant equal to unity $j_{\Lambda{\rm CDM}}=1$. However, compared to the Hubble constant and DP, the observational constraints on the value of the jerk parameter are rather weak such that various observational studies give values approximately in the range $-5\lesssim j_{0}\lesssim 10$ \cite{Sahni03,Visser04,Rapetti07,Cattoen08,Wang09,Vitagliano10,Capozziello11,Xia12}.  In recent times, there has been a great deal of interest in studying kinematics of the Universe through model independent and  phenomenological parametrizations of DP \cite{Riess04,CunhaLima08,Xu08,Xu09b,Cunha09,Nair12,Campo12}. In particular, the parametrizations $q=q_0+q_1 z$ and  $q=q_0+q_1 (1-a/a_0)$ (Chevallier-Polarski-Linder parametrization\footnote{The Chevallier-Polarski-Linder parametrization was first introduced in Ref. [38] phenomenologically for describing a dark energy source with a variable EoS parameter.}) of DP\footnote{Here $q_1$ is a constant and the subscript $_{0}$ stands for the current value of the considered parameter.}, which are linear in cosmic redshift $z$  and scale factor $a$, have been  frequently utilized in the literature to study kinematics of Universe with the available observational data \cite{Riess04,CunhaLima08,Xu08,Xu09b,Cunha09,Nair12}. In the next section, taking the DP as our starting point, we present a logical strategy that leads to these two well known parametrizations of DP and an additional new parametrization; linear parametrization of DP in cosmic time $t$, viz., $q=q_0+q_1(1-t/t_0)$.

\section{Linear parametrizations of the DP}
\label{sec:LVDPlaws}

The evolution of DP can, in principle, be described by functions of some other cosmic parameters such as cosmic time $t$, cosmic scale factor $a$, cosmic redshift $z$, which can also be re-written in terms of each other. However, we don't know the accurate function that describes evolution of DP. So we make a very reasonable assumption that the unknown function describing the DP can be approximated by its Taylor expansion in terms of $x$:
\begin{equation}
\label{eqn:Taylor1}
q(x)=q_{0}+q_{1}\left(1-\frac{x}{x_{0}}\right)+q_{2}\left(1-\frac{x}{x_{0}}\right)^{2}+...,
\end{equation}
where $x$ is a cosmological parameter to be chosen such as $z$, $a$, $t$. It is obvious that the first approximation to the unknown function $q(x)$ gives a  constant DP that corresponds to either power-law or exponential expansion, and is independent of the choice of the cosmological parameter for Taylor expansion. However, for the higher order approximations, one should choose a reasonable cosmological parameter (e.g., $z$, $a$, $t$) for the Taylor expansion and the number of terms to be retained in the Taylor expansion. The goodness of the fit to the observational data will be related with (i) the chosen parameter ($z$, $a$, $t$) for the Taylor expansion and (ii) the number of terms considered. It is obvious that, for a chosen parameter ($z$, $a$, $t$) for the Taylor expansion, one would obtain better fit to the observational data when the number of terms in Taylor expansion is increased. However, we should consider not only the goodness of fit but also the simplicity of the model (namely, number of terms in the Taylor expansion) in accordance with the Occam's razor principle. If too many terms are considered, then the allowed region in the parameter space could be so large that it would not be possible to get firm conclusions. On the other hand, goodness of the fit to the observational data can also be improved by changing the cosmic parameter considered for the Taylor expansion instead of increasing the number of terms in the expansion. Hence, we retain the first two terms in the Taylor expansion of $q(x)$, viz.,
\begin{equation}
\label{eqn:Taylor2}
q(x)=q_{0}+q_{1}\left(1-\frac{x}{x_{0}}\right),
\end{equation}
and utilize it, in what follows, to obtain the DP laws linearly varying with $z$, $a$ and $t$. It is worth noting here that this parametrization of DP carries one extra degree of freedom in comparison to the $\Lambda$CDM model. However, it is the same case with commonly considered scalar field dark energy models.

The first \textit{linearly varying DP} (LVDP), we shall consider, is the LVDP in terms of cosmic redshift $z$ (LVDP$z$), which has been the first and most commonly used law in the literature for studying the kinematics of the Universe using $H(z)$ and/or SNe Ia data \cite{Riess04,CunhaLima08,Cunha09,Nair12}. To get LVDP$z$, we simply substitute $\frac{x}{x_{0}}=\frac{a_{0}}{a}=1+z$ in (\ref{eqn:Taylor2}) and absorb the minus sign with $q_1$:
\begin{equation}
\label{eqn:LVDPz}
q=q_{0}+q_{1}z, \quad (\textnormal{LVDP}z)
\end{equation}
where $z=-1+\frac{a_{0}}{a}$ is the cosmic redshift. We note that the DP grows monotonically with no limits as we go back to earlier times of the Universe which would make the results obtained by LVDP$z$ unreliable particularly if the observations under consideration involve high redshift data. For instance, if we assume that the current value of the DP is $q_{z=0}\sim -0.5$ and the acceleration started at $z\sim 0.5$, then $q=1$ is reached too recently when $z=3$, whereas Big Bang nucleosynthesis constrains DP to be $q\sim 1$ when $z\sim 10^{9}$ \cite{Simha08}. The LVDP$z$, on the other hand, is well behaved in the future such that the Universe either expands forever (provided that $q_{0}\geq q_{1}-1$) or ends with a Big Rip in finite future (provided that $q_{0}<q_{1}-1$). However, the predicted future of the Universe using the LVDP$z$ cannot be considered so reliable since the observational constraints obtained using it with high redshift data are not reliable. 

Unlimited growth of the DP for large $z$ values in the LVDP$z$ led some authors \cite{CunhaLima08,Xu08,Xu09b,Nair12} to propose linearly varying DP in scale factor $a$ (LVDP$a$), which can be obtained here by substituting $\frac{x}{x_{0}}=\frac{a}{a_{0}}$ in (\ref{eqn:Taylor2}):
\begin{eqnarray}
\label{eqn:LVDPa}
q=q_{0}+q_{1}\left(1-\frac{a}{a_{0}}\right). \quad(\textnormal{LVDP}a)
\end{eqnarray}
It is regarded as the Chevallier-Polarski-Linder (CPL) parametrization of DP and can be written in terms of redshift as follows:
\begin{eqnarray}
q=q_0+q_1\frac{z}{1+z}.
\end{eqnarray}
At this stage, it should be noted that in all the LVDP models the transition of the Universe from deceleration to acceleration demands the constant $q_1$ to be positive. One may observe that LVDP$z$ and LVDP$a$ models exhibit similar behaviors at low redshift values (see Table \ref{table:LVDP}). On the other hand, LVDP$a$ gets finite values in the whole past of the Universe, which makes it more reliable than the LVDP$z$, particularly for observational studies involving high redshift data. Nevertheless, this time we cannot rely on the predictions about the future of the Universe since the LVDP$a$ grows very rapidly, such that the DP $q\rightarrow -\infty$ (hence the EoS parameter of the effective fluid $w\rightarrow -\infty$ when the general relativity is considered) as $z\rightarrow-1$, which we will call \textit{Super Big Rip}, and cannot be taken as a realistic prediction (One may see Ref. \cite{Odintsov05} for a detailed discussion on the properties of singularities in the presence of phantom dark energy). 

\begin{table}\centering\small
\caption{Comparison of LVDP$t$, LVDP$a$ and LVDP$z$ models.}
\begin{tabular}{lccc}
\hline\hline
 & LVDP$t$  & LVDP$a$ & LVDP$z$ \\
\hline\hline\\
$q(t)$ & $q_{0}+q_{1}(1-\frac{t}{t_0})$  & Not available & Not available \\[6pt]

$q(a)$ & $q_{0}+q_{1}\left(1-\frac{2+2q_{0}+2q_{1}}{q_{1}+(2+2q_{0}+q_{1})(a/a_{0})^{-1-q_{0}-q_{1}}}\right)$  & $q_{0}+q_{1}(1-\frac{a}{a_{0}})$ & $q_0+q_1\left(-1+\frac{a_{0}}{a}\right)$ \\[6pt]

$q(z)$ & $q_{0}+q_{1}\left(1-\frac{2+2q_{0}+2q_{1}}{q_{1}+(2+2q_{0}+q_{1})(1+z)^{1+q_{0}+q_{1}}}\right)$  & $q_0+q_1\frac{z}{1+z}$ & $q_0+q_1z$ \\[6pt]

$q(z\rightarrow\infty)$ & $q_{0}+q_{1}$  & $q_{0}+q_{1}$ & $+\infty$  \\[6pt]

 $q(z=0)$ & $q_{0}$ & $q_{0}$ & $q_{0}$    \\[6pt]

 $q(z\sim 0)$ & $\sim q_{0}+\frac{q_{1}}{2}(2+2q_{0}+q_{1})(z-\frac{2+q_{0}}{2}z^{2})$ & $\sim q_{0}+q_{1}(z-z^2)$ & $q_0+q_1z$ \\[6pt]

 $q(z=-1)$ & $-q_{0}-q_{1}-2$ &$-\infty$ & $q_{0}-q_{1}$  \\[6pt]

Future & Big Rip  & Super Big Rip & Big Rip if $q_{0}<q_{1}-1$ \\[6pt]
 &   &  & Expands forever if $q_{0}\geq q_{1}-1$ \\[6pt]
\hline\hline
\end{tabular}
\label{table:LVDP}
\end{table}

The strategy under consideration has led to LVDP$z$ and LVDP$a$ models, which are the most considered models in the literature \cite{Riess04,CunhaLima08,Xu08,Xu09b,Cunha09,Nair12} to obtain the observational constraints on the kinematics of the Universe, and correspond to Taylor expansions of the unknown function of the DP up to the second term in terms of cosmic redshift $z$ and scale factor $a$, respectively. However, the most essential concept when we are dealing with a dynamical system is time, and hence the cosmic time $t$ in cosmology. Also, the strategy we followed naturally leads to linear parametrization of the DP in terms of cosmic time $t$. In accordance with this, similar to the LVDP$z$ and LVDP$a$ models, Taylor expansion of the unknown function of the DP in terms of $t$ up to the second term gives a linearly varying DP in time $t$ (LVDP$t$), which was recently suggested by Akarsu and Dereli \cite{Akarsu12a,Akarsu12b} phenomenologically for a reasonable generalization of cosmological models constructed under the assumption of constant DP. Accordingly, LVDP$t$ can be obtained simply by substituting $\frac{x}{x_{0}}=\frac{t}{t_{0}}$ into (\ref{eqn:Taylor2}):
\begin{eqnarray}
\label{eqn:LVDPt}
q=q_{0}+q_{1}\left(1-\frac{t}{t_{0}}\right). \quad(\textnormal{LVDP}t)
\end{eqnarray}
In contrast to the LVDP$z$ and LVDP$a$ models, this model cannot be used for observational analysis directly. One should first solve it for the scale factor explicitly, and then obtain the time red-shift relation\footnote{See equation (\ref{eq19}) in Section \ref{sec:LVDPvsLCDM} for the relation between $t$ and $z$ in the LVDP$t$ model.}. Once it is done, we obtain the LVDP$t$ in terms of red-shift as follows:
\begin{equation}
q=q_{0}+q_{1}\left(1-\frac{2+2q_{0}+2q_{1}}{q_{1}+(2+2q_{0}+q_{1})(1+z)^{1+q_{0}+q_{1}}}\right).
\end{equation}
We note that at low redshift values all the three LVDP models behave similarly while only the LVDP$t$ never diverges throughout the history of the Universe, namely, it is finite both at the beginning $a\rightarrow 0$ ($z\rightarrow\infty$ ) and at the end $a\rightarrow \infty$ ($z\rightarrow-1$) of the Universe (see Table \ref{table:LVDP}). It was also shown in Ref. \cite{Akarsu12b} that it mimics the $\Lambda$CDM cosmology for a long passage of time. This is important since we know that any cosmological model that can be a good candidate should not deviate from $\Lambda$CDM kinematics a lot at least in the observational range of the Universe. We note that Universe inevitably evolves into a super-exponential expansion phase with a finite DP value less than $-1$ that imposes a Big Rip end to the Universe \cite{Caldwell03}, which is characterized by $\rho\rightarrow\infty$, $p\rightarrow\infty$ and $p/\rho\rightarrow {\rm const.}$ as $t\rightarrow t_{\rm BR}$ \cite{Odintsov05}. This also imposes, in general relativity, the presence of dark energy source that can cross below the phantom divide line ($w<-1$), which is dubbed as quintom dark energy \cite{Nesseris07,Cai}. This is a behavior that is favored persistently by many observational studies \cite{Godlowski05,Zhao07,Alberto12,Novosyadlyj12,Parkinson12,Odintso13,PlanckCosmological}. We would also like to comment that LVDP$t$ behavior has recently been obtained by Akarsu and Dereli \cite{Akarsu13a} in a cosmological model while studying the late time acceleration of the 3-space in a higher dimensional steady state Universe in dilaton gravity.

In Table \ref{table:LVDP}, one may see a comparison of the three different linear parametrizations of the DP, viz., LVDP$t$, LVDP$a$ and LVDP$z$. Arising from Taylor series expansion approach, LVDP$t$ shares the same physical, at least the mathematical, motivation with LVDP$z$ and LVDP$a$, the two commonly used linear parametrizations of DP in the literature. Interesting theoretical features of LVDP$t$ model we discussed above prompt us to study observational constraints on its parameters that we present in the next section. We also study the observational constraints on LVDP$a$ and LVDP$z$ models for the sake of comparison with the LVDP$t$ model.

\section{Observational constraints on the LVDP models from $H(z)$+SN Ia data }
\label{sec:obsLVDP}

The parameters of the LVDP models are constrained considering the 25 observational $H(z)$ data points spanning over the redshift range $0.07 < z < 1.750$ \cite{ref:Zhang2012} and 580 SNe Ia data points of Union2.1 sample \cite{ref:Suzuki2012} spanning over the redshift range $0.015 < z < 1.414$ and using the Markov Chain Monte Carlo (MCMC) method, whose code is based on the publicly available package {\bf cosmoMC} \cite{ref:MCMC}. The details of methodologies and observational data points used for obtaining the constraints on model parameters are given in Appendices \ref{sec:axHz} and \ref{sec:axSN}. We combine $H(z)$ and SNe Ia data in order to obtain tighter constraints on the model parameters and to avoid degeneracy in the observational data. Since independent cosmological probes provide the $H(z)$ and SNe Ia data, it is reasonable to define the total likelihood as the product of separate likelihoods of the two probes, viz.,
\begin{equation}\label{7}\nonumber
\chi^2_{\rm total}=\chi^2_{OHD}+\chi^2_{\rm SN}.
\end{equation}

The Hubble parameters in terms of cosmic redshift $z$ used for fitting the data and obtained from the LVDP$z$, LVDP$a$ and LVDP$t$ models, respectively, read as 
\begin{equation}
H(z)=H_{0}(1+z)^{1+q_{0}-q_{1}}e^{q_{1}z}, \quad(\textnormal{LVDP}z)
\end{equation}
\begin{equation}
H(z)=H_{0}(1+z)^{1+q_{0}+q_{1}}e^{-q_{1}\frac{z}{1+z}}, \quad(\textnormal{LVDP}a)
\end{equation}
\begin{equation}
H(z)=\frac{H_{0}}{4(1+q_0+q_1)^2 } [q_1(1+z)^{-(1+q_0+q_1)/2}+ (2+2q_0+q_1)(1+z)^{(1+q_0+q_1)/2}]^2. \quad(\textnormal{LVDP}t)
\end{equation}

The 1D marginalized distribution on individual parameters and 2D contours with 68.3 \%, 95.4 \% and 99.73 \% confidence limits are shown in Fig.\ref{fig:LVDPtcontour} for the LVDP$t$ model, in Fig.\ref{fig:LVDPzcontour} for the LVDP$z$ model and in Fig.\ref{fig:LVDPacontour} for the LVDP$a$ model.
\begin{figure}[htb!]
     \begin{center}
        \subfigure[LVDP$t$]{%
            \label{fig:LVDPtcontour}
            \includegraphics[width=0.31\textwidth]{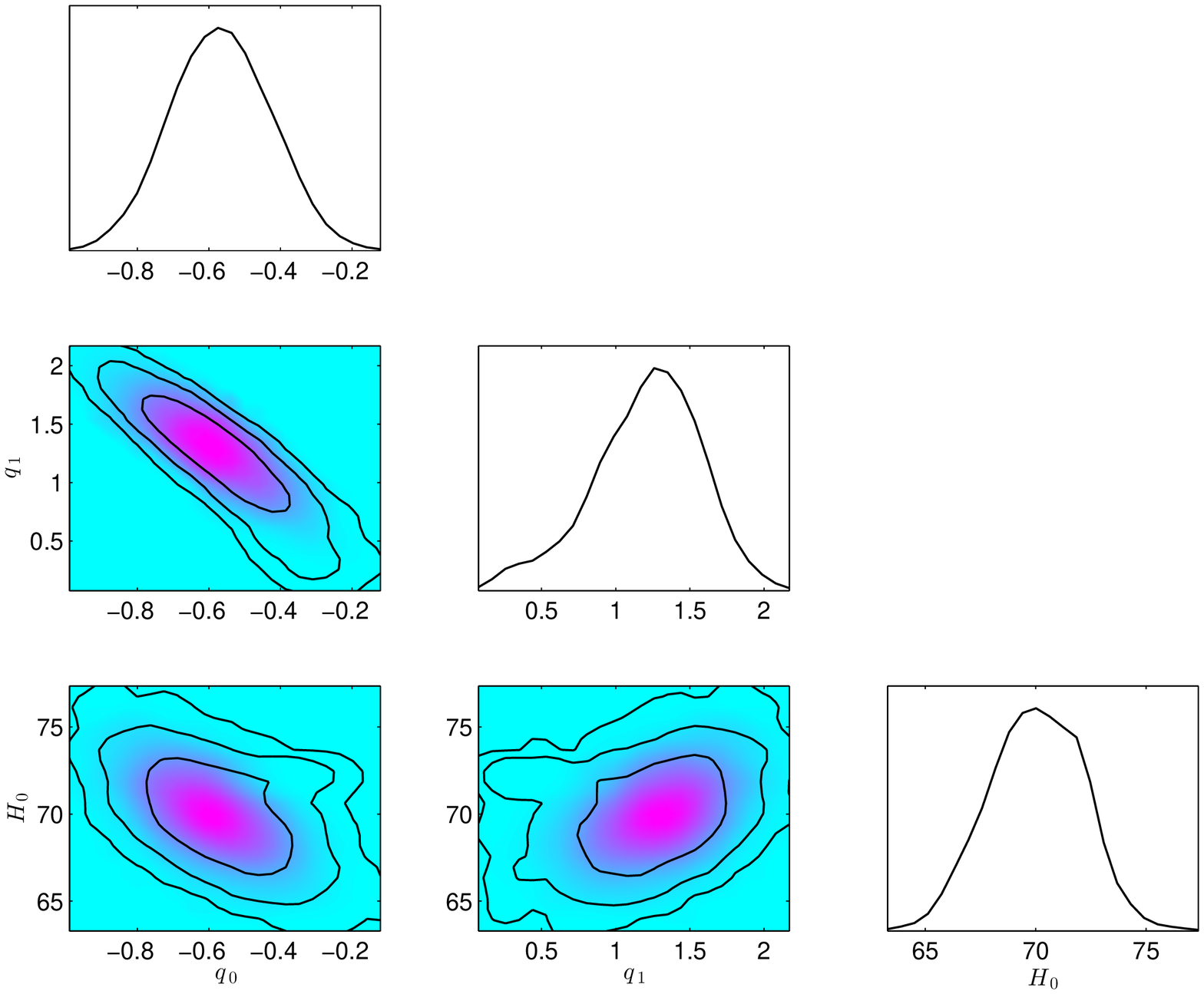}
        }%
        \subfigure[LVDP$z$]{%
           \label{fig:LVDPzcontour}
           \includegraphics[width=0.32\textwidth]{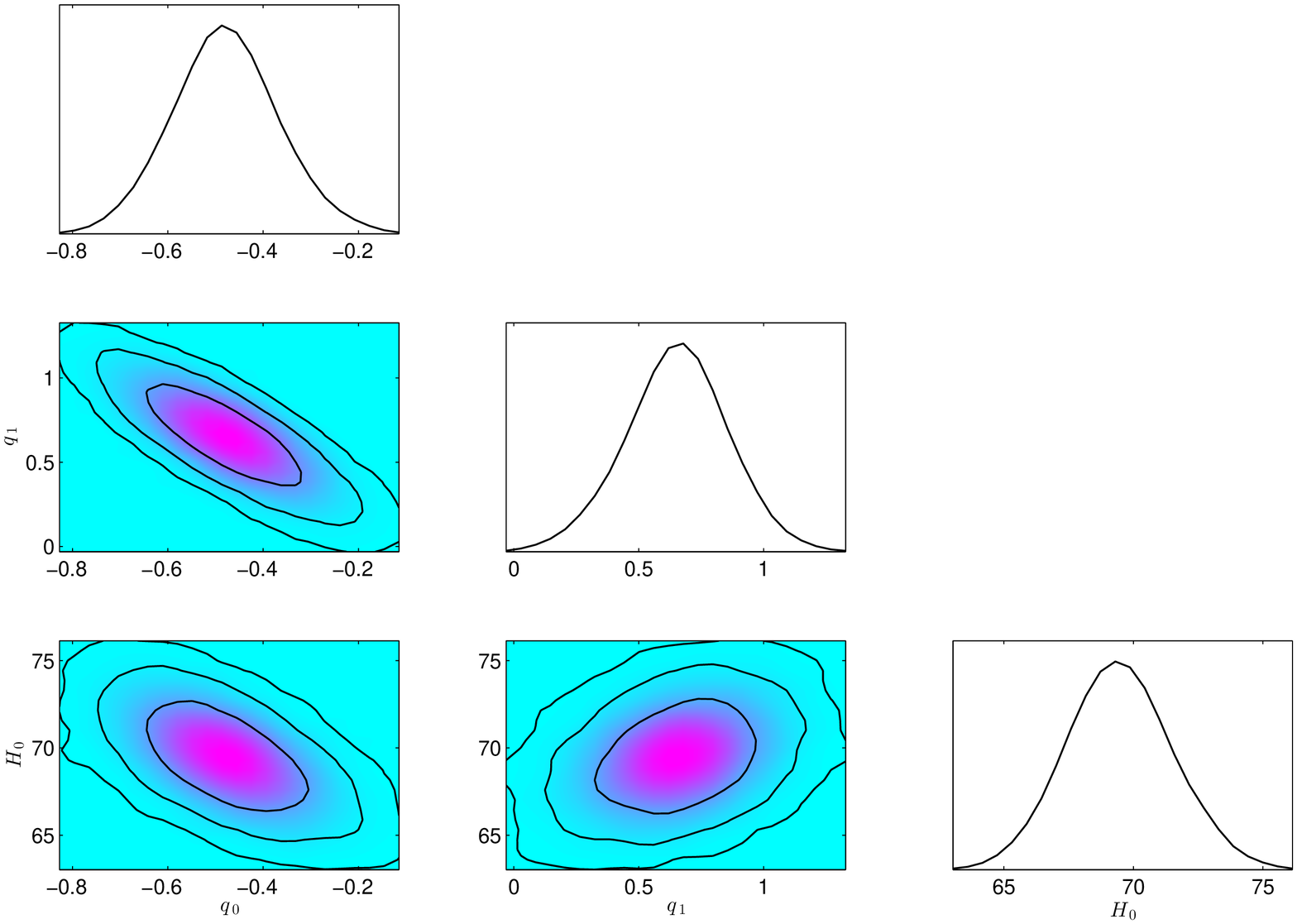}
        }
        \subfigure[LVDP$a$]{%
            \label{fig:LVDPacontour}
            \includegraphics[width=0.32\textwidth]{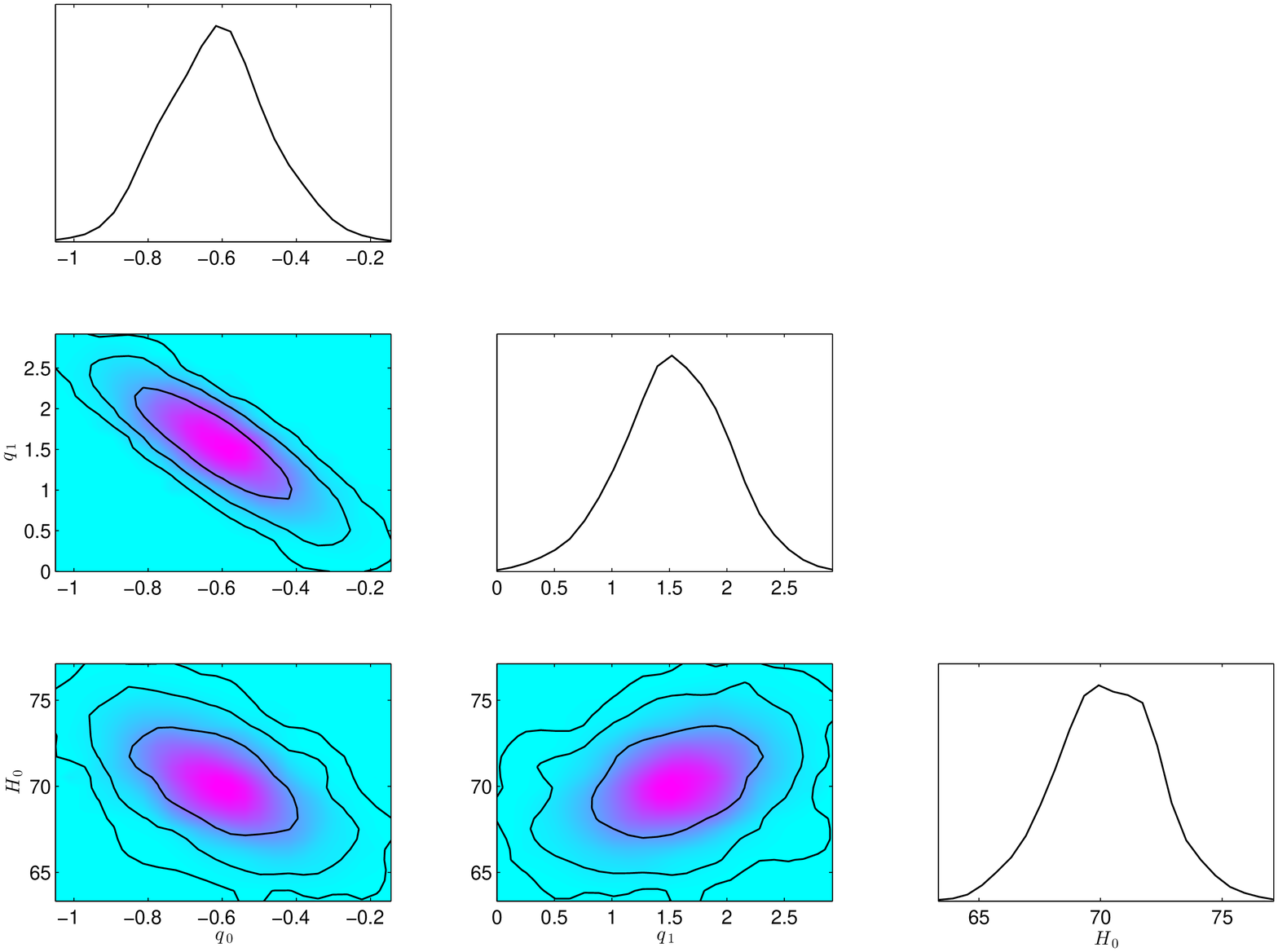}
        }%
        
    \end{center}
    \caption{%
        The 1D marginalized distribution on individual parameters of (a) LVDP$t$, (b) LVDP$z$ and (c) LVDP$a$ and 2D contours with 68.3 \%, 95.4 \% and 99.7 \% confidence levels are obtained by using $H(z)$+SN Ia data points. The shaded regions show the mean likelihood of the samples.
     }%
	\label{fig:contours}
\end{figure}

The mean values of the model parameters $q_{0}$, $q_{1}$ and the Hubble constant $H_{0}$ of the LVDP$t$, LVDP$a$ and LVDP$z$ models constrained with $H(z)$+SN Ia data are given in Table \ref{table:LVDPobs}. The $1\sigma$ errors, $\chi^2_{min}$, $\chi^2_{min}$/dof and goodness of fit (GoF) are also given in Table \ref{table:LVDPobs}.
\begin{table}\centering\small
\caption{Mean values with $1\sigma$ errors of the parameters of LVDP$t$, LVDP$z$ and LVDP$a$ models constrained with $H(z)$+SN Ia data. $\chi^2_{min}$/dof and GoF values are also given.} 
\begin{tabular}{llll}
\hline\hline Parameters & LVDP$t$ & LVDP$a$  & LVDP$z$  \\ \hline\hline\\
$q_0$ & $-0.565_{-0.135}^{+0.141}$ & $-0.613_{-0.141}^{+0.137}$ & $-0.478_{-0.107}^{+0.106}$\\[6pt]
$q_1$ &  $1.206_{-0.344}^{+0.352}$ & $1.540_{-0.442}^{+0.453}$ & $0.652_{-0.197}^{+0.194}$ \\[6pt]
$H_0$ ($\frac{\rm km}{\rm s \cdot Mpc}$) & $70.016_{-2.096}^{+2.089}$ & $70.260_{-2.102}^{+1.887}$ & $69.470_{-1.964}^{+1.991}$ \\[6pt]
$\chi^2_{min}$ & $556.489$ & $556.508 $ & $557.314$ \\[6pt]
 $\chi^2_{min}$/dof & $0.91981$ & $0.91984$ & $0.92118$ \\[6pt]
GoF & $92.137$  & $92.129$ & $91.763$ \\[6pt]
\hline\hline
\end{tabular}
\label{table:LVDPobs}
\end{table}
We note that the LVDP$t$ model fits the observational data better than the LVDP$a$ and LVDP$z$ models. LVDP$a$ model, on the other hand, fits the observational data better than the LVDP$z$ model.

In Table \ref{table:cospar}, we give the values of the cosmological parameters that we obtained using the three LVDP models separately; age of the present Universe $t_{0}$;  Hubble constant $H_{0}$, current value of the DP $q_{0}$ and jerk parameter $j_{0}$, time passed since the accelerating expansion started $t_{0}-t_{\rm tr}$ and redshift of the onset of the accelerating expansion $z_{\rm tr}$.
\begin{table}\centering\small
\caption{Mean values with $1\sigma$ errors of some important cosmological parameters pertaining to LVDP$t$, LVDP$z$ and LVDP$a$ models.} 
\begin{tabular}{llll}
\hline\hline Parameters & LVDP$t$ & LVDP$a$  & LVDP$z$  \\ \hline\hline\\

$t_0 $ (Gyr)& $13.460\pm 0.899$ & $13.138 \pm  0.754$ & $11.934\pm 0.535$ \\[6pt]
$H_0$ ($\rm km\,\rm s^{-1}\,\rm Mpc^{-1}$) & $70.016_{-2.096}^{+2.089}$ & $70.260_{-2.102}^{+1.887}$ & $69.470_{-1.964}^{+1.991}$ \\[6pt]
$q_0$ & $-0.565_{-0.135}^{+0.141}$ & $-0.613_{-0.141}^{+0.137}$ & $-0.478_{-0.107}^{+0.106}$\\[6pt]
$j_{0}$ & $1.326\pm 0.599$ & $1.680\pm 0.645$ & $0.631\pm 0.290$ \\[6pt]
 $t_{0}-t_{\rm tr}$ (Gyr) & $6.312\pm1.288$ & $6.082\pm1.066$ & $6.504\pm0.752$ \\[6pt]
 $z_{\rm tr}$ & $0.703_{-0.148}^{+0.356}$  & $0.662_{-0.107}^{+0.198}$ & $0.733_{-0.095}^{+0.148}$ \\[6pt]
\hline\hline
\end{tabular}
\label{table:cospar}
\end{table}

We note that all the values obtained using LVDP$t$ and LVDP$a$ models are consistent within $1\sigma$ error region. The Hubble constant values are consistent within $1\sigma$ error region in the three LVDP models. However, the age of the present Universe obtained in LVDP$z$ model has a too low value ($11.934\pm 0.535$ Gyr) and is not consistent within $1\sigma$ error region with the ones obtained in LVDP$t$ and LVDP$a$ models. The current values of the DP obtained in all three LVDP models are consistent within $1\sigma$ error region. The current values of the jerk parameter obtained in LVDP$t$ and LVDP$a$ models are consistent within $1\sigma$ error region and also cover $j_{\Lambda{\rm CDM}}=1$. However, the current value of the jerk parameter obtained using LVDP$z$ model is not only inconsistent with the ones obtained in LVDP$t$ and LVDP$a$ models but also does not cover $j_{\Lambda{\rm CDM}}=1$ within $1\sigma$ error region. Considering its poor fit to the data and too low age for the current Universe, we conclude that the LVDP$z$ model is less reliable than the  LVDP$t$ and LVDP$a$ models for obtaining the kinematics of the Universe. According to LVDP$t$ and LVDP$a$ models, the accelerating expansion of the Universe starts when the Universe was $\sim 7$ Gyr old ($\sim 6$ Gyr ago from now) and at redshift $\sim 0.7$.

Jerk parameter is a key parameter to investigate the deviations from the $\Lambda$CDM model in which jerk parameter is simply a constant equal to $1$. We note that the present value of the jerk parameter does not deviate from $j=1$ a lot in all the three LVDP models but it is covered only by the LVDP$t$ model within $1\sigma$ error region. 

\section{The dynamics of the LVDP$t$ model}
\label{sec:LVDPvsLCDM}

In this section, we discuss the dynamics of LVDP$t$ model taking into account the effective fluid responsible for the LVDP$t$ kinematics within the framework of general relativity in comparison dynamics of the $\Lambda$CDM model (considering effective fluid only, which consists of the conventional vacuum energy and dust).

The scale factor in terms of cosmic time $t$ for the LVDP$t$ model (\ref{eqn:LVDPt}) is as follows:
\begin{equation}
\label{eqn:LVDPtSF}
a(t)=a_{1} e^{\frac{2}{1+q_0+q_1}\; \tanh^{-1}\left[\frac{q_1t}{(1+q_0+q_1) t_{0}}-1\right]},
\end{equation}
where $a_{1}$ is a constant of integration. The Hubble parameter
\begin{equation}
H=\frac{2t_{0}}{t[2(1+q_0+q_1)t_0-q_1t]},
\end{equation}
and the dimensionless jerk parameter
\begin{equation}
j =\frac{3 q_1^2 t^2}{2 t_0^2}-\frac{3 q_1  (q_0+q_1)t}{t_0}+(q_0+q_1) (2 q_0+2 q_1+1).
\end{equation}
In general relativity within the framework of spatially flat FRW spacetime the effective energy density $\rho=(3/8\pi G)H^2$ and pressure $p=-(c^2/8\pi G)(2\dot{H}+3H^2)$. Using these we obtain the energy density and pressure of the effective fluid that would give rise to the LVDP$t$ kinematics in general relativity as
\begin{equation}
\label{eqn:rho}
\rho=\frac{3t_{0}^2}{2\pi Gt^2[2(1+q_0+q_1)t_0-q_1t]^2},
\end{equation}
\begin{equation}
\label{eqn:p}
p=\frac{c^2t_{0}[-2q_1t+(2q_0+2q_1-1)t_{0}]}{2\pi Gt^2[2(1+q_0+q_1)t_0-q_1t]^2},
\end{equation}
respectively. Consequently, we obtain the effective EoS parameter as follows:
\begin{equation}
\label{eqn:equation of state}
w=-1+\frac{2(1+q_0+q_1)}{3}-\frac{2q_1t}{3t_{0}}.
\end{equation}
Further, we would like to note here that one is able to re-write all these cosmological parameters that are given in terms of cosmic time $t$ in terms cosmic redshift $z$ using the relation
\begin{equation}\label{eq19}
t=\frac{2(1+q_0+q_1)t_{0}}{q_1+(2+2q_0+q_1)(1+z)^{1+q_0+q_1}},
\end{equation}
between the $t$ and $z$ that can be obtained using $a=1/(1+z)$ with the choice $a(t_{0})=1$ for the present size of the Universe.

We see that the parameters $H>0$, $\rho$ and $p$ diverge at $t=0$ and $t=2(1+q_0+q_1)t_0/q_1$ while $q$, $j$ and $w$ are finite at these two moments. This in turn implies that the Universe governed by LVDP$t$ begins with a Big Bang at $t=0$ and ends in a Big Rip at $t_{\rm BR}=2(1+q_0+q_1)t_0/q_1$. The de Sitter time that corresponds to $q=-1$ reads as $ t_{\rm ds}=(1+q_0+q_1)t_0/q_1$. It is interesting to observe the coincidence that the de Sitter time in the LVDP$t$ model is exactly half the Big Rip time, that is, $ t_{\rm ds}=\frac{1}{2}t_{\rm BR}$.

In Table \ref{table:LCDMobs}, we give the mean values with $1\sigma$ errors of the parameters of LVDP$t$ and $\Lambda$CDM models constrained with $H(z)$+SN Ia data as in the previous section. The values of $\chi^2_{min}$, $\chi^2_{min}$/dof and GoF are also displayed in Table \ref{table:LCDMobs}. We note that both the models have almost the same GoF to the observational data. However, it should be noted that the LVDP$t$ model carries one extra degree of freedom, as in the simplest scalar field models of dark energy, compared to the $\Lambda$CDM model. This also causes the errors in the LVDP$t$ law to be larger than the ones in the $\Lambda$CDM model.
\begin{table}\centering\small
\caption{Mean values with $1\sigma$ errors of the parameters of LVDP$t$ and $\Lambda$CDM  models constrained with $H(z)$+SN Ia data. $\chi^2_{min}$/dof and GoF values are also displayed.} 
\begin{tabular}{llll}
\hline\hline Parameters & LVDP$t$ & $\Lambda$CDM  \\ \hline\hline\\
$q_0$ & $-0.565_{-0.135}^{+0.141}$ &  $-0.556\pm 0.046$ \\[6pt]
$q_1$ &  $1.206_{-0.344}^{+0.352}$ & --  \\[6pt]
$H_0$ (${\rm km}\,s^{-1}\,{\rm Mpc}^{-1}$) & $70.016_{-2.096}^{+2.089}$ & $70.697_{-2.020}^{+1.667}$ \\[6pt]
$\chi^2_{min}$ & $556.489$ & $556.499$  \\[6pt]
 $\chi^2_{min}$/dof & $0.91981$ & $0.91983$ \\[6pt]
GoF & $92.137$  & $92.133 $ \\[6pt]
\hline\hline
\end{tabular}
\label{table:LCDMobs}
\end{table}

We give cosmological parameters including the present values of the effective energy densities $\rho_{0}$, pressures $p_{0}$ and EoS parameters $w_{0}$ for both models in Table \ref{table:LVDPLCDM}.
\begin{table}\centering\small
\caption{Mean values with $1\sigma$ errors of some important cosmological parameters  related to LVDP$t$ and $\Lambda$CDM models.} 
\begin{tabular}{llll}
\hline\hline Parameters & LVDP$t$ & $\Lambda$CDM \\ \hline\hline\\
$t_0 $ (Gyr)& $13.460\pm 0.899$ & $13.389 \pm  0.289$ \\[6pt]
$H_0$ (${\rm km}\,s^{-1}\,{\rm Mpc}^{-1}$) & $70.016_{-2.096}^{+2.089}$ & $70.697_{-2.020}^{+1.667}$ \\[6pt]
$q_0$ & $-0.565_{-0.135}^{+0.141}$ & $-0.556\pm 0.046$  \\[6pt]
$j_{0}$ & $1.326\pm 0.599$ & $1$ \\[6pt]
 $t_{0}-t_{\rm tr}$ (Gyr) & $6.312\pm1.288$ & $6.156\pm 0.366$  \\[6pt]
 $z_{\rm tr}$ & $0.703_{-0.148}^{+0.356}$  & $0.682\pm 0.082$ \\[6pt]
 $\rho_{0}\;(10^{-27}$ kg m$^{-3}$) & $9.209\pm 0.545$ & $9.389\pm 0.481$ \\[6pt]
$p_{0}\;(10^{-10}$ Pa) & $-5.890\pm 0.976$ & $-5.952\pm 0.520$ \\[6pt]
$w_{0}$ & $-0.710\pm 0.090$ & $-0.704\pm 0.030$\\
\hline\hline
\end{tabular}
\label{table:LVDPLCDM}
\end{table}
We observe that the present values of the cosmological parameters as well as the parameters related with the onset of accelerating expansion obtained in LVDP$t$ and $\Lambda$CDM models are indistinguishable. Strictly speaking all the values obtained using LVDP$t$ and $\Lambda$CDM are consistent even at $1\sigma$ level. In particular, the values we obtained for the Hubble constant from the latest $H(z)$+SNI a Union2.1 compilation are completely consistent even within $1\sigma$ error region; $H_0= 70.016_{-2.096}^{+2.089}\,{\rm km}\,s^{-1}\,{\rm Mpc}^{-1}$ in the LVDP$t$ model and $H_0= 70.697_{-2.020}^{+1.667}\,{\rm km}\,s^{-1}\,{\rm Mpc}^{-1}$ in the $\Lambda$CDM model. These values are consistent with the low value $H_0= 70.0\pm 2.2\,{\rm km}\,s^{-1}\,{\rm Mpc}^{-1}$ that has been found in the recent WMAP-9 analysis \cite{Bennett12} assuming the base six-parameter $\Lambda$CDM model. On the other hand, only the one obtained in LVDP$t$ model is consistent with the low value $H_{0}=67.3\pm 1.2\,{\rm km}\,s^{-1}\,{\rm Mpc}^{-1}$ from the Planck+WP+highL analysis within $1\sigma$ error region in the \textit{Planck} experiment\cite{PlanckCosmological}.

We expect these two models to differ slightly at earlier times of the Universe and differ significantly in the far future. Therefore, we compare the asymptotic values of various parameters of the LVDP$t$ and $\Lambda$CDM models in Table \ref{table:extremes}.
\begin{table}\centering\footnotesize
\caption{Mean values and asymptotic limits with $1\sigma$ errors of various parameters pertaining to the LVDP$t$ and $\Lambda$CDM models.}
\centering
\begin{tabular}{|c|c|c|c|c|c|c|c|c|c|}
\hline
Model $\rightarrow$ & \multicolumn{3}{c|}{LVDP$t$} &  \multicolumn{3}{c|}{$\Lambda$CDM}\\\hline
Parameter & $z\rightarrow \infty$ & $z=0$ & $z\rightarrow -1$ & $z\rightarrow \infty$ & $z=0$&$z\rightarrow -1$\\\hline
$H\;$(km s$^{-1}$ Mpc$^{-1}$) & $\infty$ & $70.016_{-2.096}^{+2.089}$ & $\infty$ & $\infty$ & $70.697_{-2.020}^{+1.667}$ & $59.336\pm 2.592$ \\[6pt]
$q$ & $0.640\pm 0.238$& $-0.565_{-0.135}^{+0.141}$ & $-2.640\pm 0.238$& $0.5$ & $-0.556\pm 0.046$ & $-1$\\[6pt]
$j$ & $1.462\pm 0.848$ & $1.326\pm 0.599$ & $11.308\pm 2.277$ & $1$& $1$ &$1$\\[6pt]
$\rho\;(10^{-27}$ kg m$^{-3}$) & $\infty$ & $9.209\pm 0.545$ & $\infty$ & $\infty$ & $9.389\pm 0.481$ & $6.614\pm 0.577$\\[6pt]
$p\;(10^{-10}$ Pa) & $\infty$ & $-5.890\pm 0.976$ & $-\infty$ & $-5.952\pm 0.520$ & $-5.952\pm 0.520$ & $-5.952\pm 0.520$ \\[6pt]
$w$ & $0.093\pm 0.158$ & $-0.710\pm 0.090$ & $-2.093\pm 0.158$ & $0$ & $-0.704\pm 0.030$ & $-1$ \\[6pt]
\hline
\end{tabular}
\label{table:extremes}
\end{table}
 We note that LVDP$t$ and $\Lambda$CDM models differ slightly (such that the results are still consistent within $1\sigma$) at the limit $z\rightarrow\infty$. In the future, on the other hand, these two models behave totally differently. While the $\Lambda$CDM model approaches continuously to the de Sitter expansion without an end, LVDP$t$ model evolves to super-exponential expansion rate with a value $q=-2.640\pm0.238$ as $z\rightarrow -1$ which indicates that the size of the Universe will diverge in finite time. The $\Lambda$CDM model is a model that can describe the evolution of the observed Universe starting from the matter dominance $w\sim 0$ (which gives $q\sim \frac{1}{2}$ in general relativity) at $z\sim 3400$. The LVDP$t$ model can be interpreted, in a similar manner, as a model that can describe the evolution of the Universe starting from the matter dominance in the context of dark energy concept, since the value $q=\frac{1}{2}$ of the $\Lambda$CDM as $z\rightarrow \infty$ is covered by this model. However, LVDP$t$ model may be describing the evolution of the Universe starting from even earlier times of the Universe than the $\Lambda$CDM model starts to describe. In the actual Universe, the physical ingredient of the Universe at redshift values $z \gtrsim 3400$ is expected to be described with an EoS parameter $w\sim\frac{1}{3}$, which gives $q\sim 1$ in general relativity, since radiation would start to be dominant during that epoch as we go back to earlier times of the Universe. We note that these values are almost covered at $1\sigma$ level by the predicted values for $z\rightarrow \infty$ in the LVDP$t$ model. This may be suggesting that the LVDP$t$ might be a good approximation for describing the evolution of the Universe starting from the Big Bang Nucleosynthesis times. However, this might be attributed the large error values on the parameters in the LVDP$t$ model.

We would like to conclude this section by comparing the continuous evolution of these models instead of some particular times of the Universe. A very useful way of comparing and distinguishing different cosmological models that have similar kinematics is to plot the evolution trajectories of the $\{q,j\}$ and $\{j,s\}$ pairs. Here $q$ and $j$ have the usual meaning and $s$ is a parameter we defined as
\begin{equation}
\label{eqn:LVDPtSFP}
s=\frac{j-1}{3(q-1)}
\end{equation}
inspired by Sahni et al.\cite{Sahni03}. Using (\ref{eqn:LVDPtSF}) in (\ref{eqn:LVDPtSFP}) we get
\begin{eqnarray}
s=\frac{3 q_1^2 t^2-6 t_0 q_1 (q_0+q_1) t+2 t_0^2(1+q_0+q_1) (-1+2 q_0+2 q_1) }{6 t_0 [-q_1 t+(-1+q_0+q_1)t_0]}
\end{eqnarray}
for the LVDP$t$ model. Note that the parameter $s$ we defined is slightly different from the one given as $s=\frac{j-1}{3(q-\frac{1}{2})}$ by Sahni et al.\cite{Sahni03}. We use $1$ in the place of $\frac{1}{2}$ in the original definition to avoid the divergence of the parameter $s$ when model passes through $q=\frac{1}{2}$. The parameter $s$ was originally introduced to characterize the properties of dark energy, and hence the evolution of the Universe was considered starting from pressure-less matter dominated era in general relativity which gives $q=\frac{1}{2}$ and $j=1$. However, in accordance with the LVDP$t$, here we are also interested in the possibility of describing the Universe starting from times even earlier than the pressure-less matter dominated era.

We plot evolution trajectories of the LVDP$t$ and $\Lambda$CDM models in the $j-q$ plane in Fig. \ref{fig:rq} and in the $j-s$ plane in Fig. \ref{fig:rs}, using the mean values of the model parameters given in Table \ref{table:LCDMobs} from observations. For comparison, we include also some alternatives to the $\Lambda$CDM model such as the Galileon, Chaplygin Gas and DGP models (for these models see Sami et al.\cite{Sami12} and references therein) in the figures. The arrows on the curves show the direction of evolution and the dots on the curves represent the present values of the corresponding $\{q,j\}$ and $\{j,s\}$ pairs while the black dots show the matter dominated phases of the models.

We observe that all the models have different evolution trajectories but the values of $q$, $j$ and $s$ do not deviate a lot in different models in the past of the Universe as well as in the near future of the Universe. However, the mean value of the DP parameter gets values higher than $0.5$ in the early Universe only in the  LVDP$t$ model, which may be indicating that LVDP$t$ can probe the earlier times of the Universe compared to the $\Lambda$CDM, DGP, Chaplygin gas and Galileon models. All these models evolve to the de Sitter expansion continuously while the LVDP$t$ model crosses the de Sitter line and reaches the super-exponential expansion rate at some point.

\begin{figure}
     \begin{center}
        \subfigure[]{%
            \label{fig:rq}
              \psfrag{r1}[b][b]{$j$}
				\psfrag{q3}[b][b]{$q$}
            \includegraphics[width=0.45\textwidth]{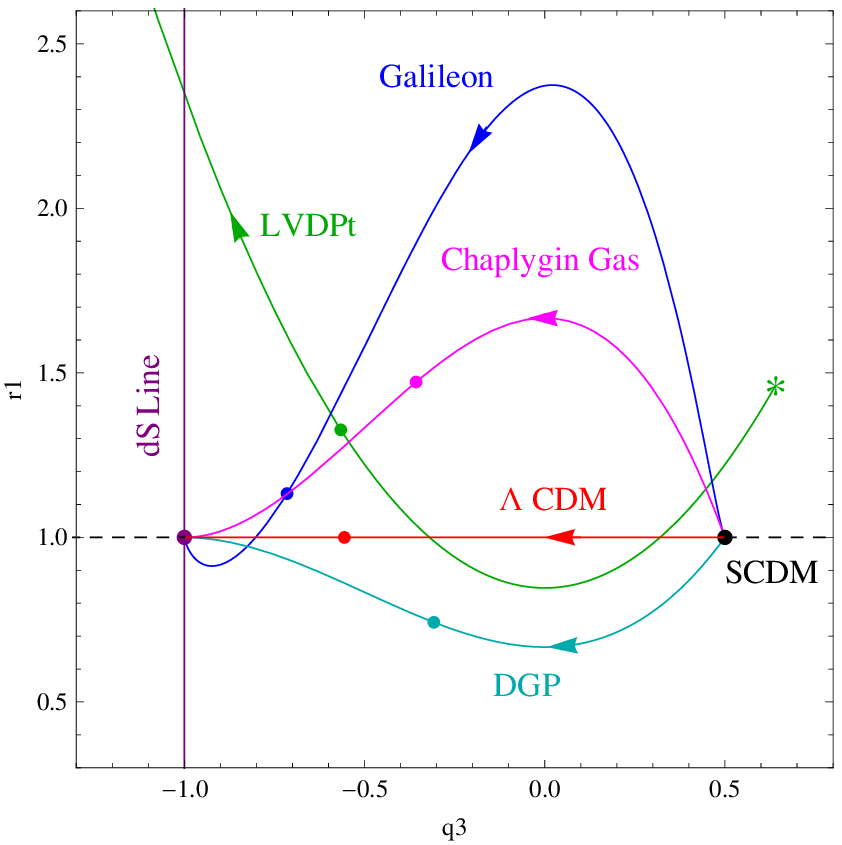}
        }%
        \subfigure[]{%
           \label{fig:rs}
           \psfrag{r1}[b][b]{$j$}
				\psfrag{s1}[b][b]{$s$}
           \includegraphics[width=0.45\textwidth]{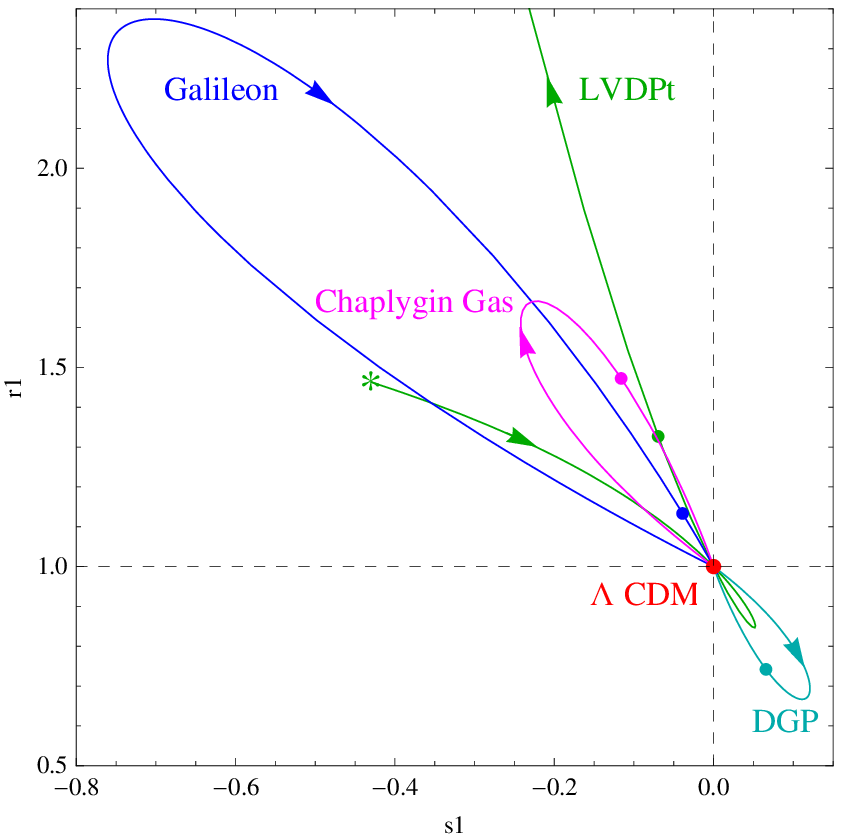}
        }

    \end{center}
    \caption{%
        \textbf{(a)} Variation of $q$ versus $j$. Vertical Purple line stands for the de Sitter (dS) state $q=-1$. \textbf{(b)} Variation of $s$ versus $j$.  Horizontal and vertical dashed lines intersect at the $\Lambda$CDM point $(0,1)$. In both panels, the Green curve corresponds to the LVDPt model. The star symbols represent the phase relatively earlier to the matter dominated phase in the LVDPt model. Thereafter, LVDPt $s-j$ curve crosses the $\Lambda$CDM statefinder point $(0,1)$ two times. The final part of the LVDPt $s-j$ curve not shown in panel (b) extends up to the Big Rip point $(-0.943,11.308)$. Similarly, LVDPt $q-j$ curve in panel (a) crosses the dS line and eventually goes up the Big Rip phase point $(-2.64,11.308)$.  Red, Cyan, Magenta and Blue curves correspond to $\Lambda$CDM, DGP, Chaplygin gas and Galileon models respectively. The arrows on the curves show the direction of evolution. The dots on the curves represent the present values of the corresponding $(s,j)$ or $(q,j)$ pair while the black dots show the matter dominated phases of the models.
     }%
	\label{fig:statefinders}
\end{figure}

\section{The history and the future of the Universe}
\label{sec:history}

In this final section, we discuss some important moments in the expansion history and the possible fates of the Universe in different models; the transition time and redshift from decelerating expansion to the accelerating expansion, the present age of the Universe, the time and redshift when the de Sitter expansion ($q=-1$) will be reached (provided it is reached) and the time when the size of the Universe will become infinite, $a\rightarrow\infty$, i.e., when $z\rightarrow -1$ (whether in infinite time or in finite time). We give these important moments in the history of the Universe in the LVDP$t$, LVDP$a$ and LVDP$z$ models in comparison with the $\Lambda$CDM model in Table \ref{tab:history}.
\begin{table*}\centering\small
\caption{Some key moments in the evolution of the Universe in the LVDP$t$, LVDP$a$ and LVDP$z$ models in comparison with the $\Lambda$CDM using the $H(z)$+SN Ia data.} 
\begin{tabular}{lcccc}
\hline\hline Parameters & LVDP$t$ & LVDP$a$  & LVDP$z$ & $\Lambda$CDM   \\ \hline\hline\\
Transition time (Gyr) & $7.148\pm 0.923$  &  $7.056\pm 0.754$ & $5.430\pm 0.529$ & $7.233\pm 0.225$ \\[6pt]
Transition redshift &  $0.703_{-0.148}^{+0.356}$  & $0.662_{-0.107}^{+0.198}$ & $0.733_{-0.095}^{+0.148}$ & $0.682\pm 0.082$  \\[6pt]
Present age (Gyr) & $13.460\pm 0.899$ & $13.138\pm 0.754$ & $11.934\pm 0.535$ & $13.389\pm 0.289$ \\[6pt]
dS time (Gyr) & $18.301\pm 1.054$ & $16.344\pm 0.839$ & $40.311$ (may be) & $\infty$  \\[6pt]
dS redshift & $-0.281_{-0.168}^{+0.120}$ & $-0.200_{-0.122}^{+0.088}$ & $-0.799$ (may be) & $-1$ \\[6pt]
Big Rip time (Gyr) & $36.602\pm 7.930$  & $27.434\pm4.015$ & $205.432$ (may be) & No \\[6pt]
\hline\hline
\end{tabular}
\label{tab:history}
\end{table*}
We note that the age of the Universe and the redshift value when the accelerating expansion started and the age of the present Universe are consistent in LVDP$t$, LVDP$a$ and $\Lambda$CDM models, though all these predict quite different futures for the Universe.

It may be noted that the LVDP$z$ model is the only unbiased model about the future of the Universe. As can be seen from Table \ref{table:LVDP}, the Universe reaches the infinitely large sizes $a\rightarrow\infty$ ($z\rightarrow -1$) in finite time or infinite time depending on the values of the parameters in LVDP$z$ model. On the other hand, irrespective of the values of parameters, the Universe takes infinite time in $\Lambda$CDM model, and finite time in the other two LVDP models for achieving infinite large sizes. Hence, although we found out above that the LVDP$z$ model is less reliable compared to the other models under consideration, it is worth first checking out the future of the Universe in LVDP$z$ model since it is the one that can give us an idea by alone whether the observational data favor a finite future or not. Accordingly, we plot the $\{q_{0},q_{1}\}$ plane with 68.3 \%, 95.4 \% and 99.7 \% confidence contours, which is divided into two regions by the de Sitter expansion line ($q_{0}-q_{1}=-1$) in Fig. \ref{fig:LVDPzBR}.
\begin{figure}\centering
\psfrag{q0}[b][b]{$t$}
\includegraphics[width=0.6\textwidth]{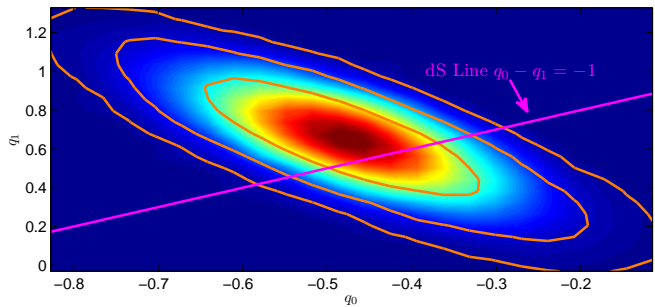}
\caption{2D contours with 68.3 \%, 95.4 \% and 99.7 \% confidence levels are shown for $q_0$ and $q_1$ in the LVDP$z$ model.The line $q_0-q_1=-1$  across the contours corresponds to the de Sitter phase ($z=-1$) in the LVDP$z$ model. The $\{q_{0},q_{1}\}$ pairs above the de Sitter line lead to Big Rip end of the LVDP$z$ Universe.}
\label{fig:LVDPzBR}
\end{figure}
One may note that the $\{q_{0},q_{1}\}$ pairs above the de Sitter line lead to a Big Rip end for the Universe. We note that within $1\sigma$ error region, the LVDP$z$ model excludes neither the forever expanding Universe ($q_{z=-1}\geq-1$) nor a Universe with a finite lifetime $q_{z=-1}<-1$. Considering the mean value $q_{z=-1}=-1.130<-1$, on the other hand, we see that it favors a Big Rip end for the Universe. We calculate numerically and give in Table \ref{tab:history} the time and redshift when the de Sitter expansion will be reached as well as when Big Rip will happen considering the mean values of the  LVDP$z$ parameters given in Table \ref{table:LVDPobs}.

We leave the LVDP$a$ out of our discussion on the future of the Universe, since in this model $q\rightarrow -\infty$ as $a\rightarrow\infty$, which does not seem realistic. We, however, calculate numerically and give in Table \ref{tab:history}, the time and  the redshift when the de Sitter expansion rate is reached and the time when the Universe will reach infinitely large sizes in this model. We are not able to make a discussion on the future of the Universe using the $\Lambda$CDM alone since it gives us no option other than a continuous approach to the de Sitter expansion that will never end. We see that  LVDP$z$ model (the unbiased LVDP model) favors a Big Rip future in line with the LVDP$t$ model. On the other hand, thinking of $\Lambda$CDM model fits the observational data better than the LVDP$z$ model, one may tend to take the future predicted by the $\Lambda$CDM model as more reliable. However, we would like to remind that the LVDP$t$ model fits the observational data well in comparison to other models under consideration, and predicts an inevitable Big Rip end for the Universe. We calculate and give the time and redshift when the de Sitter expansion is reached in LVDP$t$ and when Big Rip will happen in Table \ref{tab:history}. According to LVDP$t$ model, the Universe will start to expand with super-exponential expansion rates $\sim 5$ billions years after the present time and will reach infinitely large sizes when the Universe will be $\sim 36$ billions years old. We further discuss below the future events as the Big Rip moment approaches in an Universe described by LVDP$t$ model.

Considering phantom energy dominated Universe with EoS $w=-3/2$, Caldwell et al.\cite{Caldwell03} studied history and future of the Universe that ends with a Big Rip. On the other hand, recently Li et al.\cite{Li12fate} explored the Big Rip fate of the Universe by using a divergence-free parametrization for the EoS parameter of dark energy. Here, we discuss the future events in the Universe that obeys the LVDP$t$ model assuming general relativity to be the true theory of gravitation. In general relativity, if the Universe is expanding with a DP value less than $-1$, then there should be a dark energy source with an EoS parameter less than $-1$ in the Universe whose energy density will grow so that it eventually begins to strip the bound objects in the Universe. According to general relativity, the source for the gravitational potential is the volume integral of $\rho c^2+3p$. On the other hand,  for a gravitationally bound system with mass $M$ and radius $R$, the period of a circular orbit around this system at radius $R$ will be $P = 2\pi(R^3/GM)^{1/2}$, where $G$ is the Newton's constant.  So, as pointed out by Caldwell et al.\cite{Caldwell03}, this system will become unbound roughly when $-\frac{4\pi}{3}(\rho + 3p/c^2)R^3 = M$ that gives $-(1 + 3 w)H^2 = \frac{8\pi^2 }{P^2}$. Since the density of dark energy exceeds the energy density of the object and the repulsive gravity of the phantom energy overcomes the forces holding the object together. Accordingly, we calculate and give the events in the future of the Universe considering the best fit values of the LVDP$t$ parameters given in the Table \ref{tab:bigripresults}.
\begin{table}\centering \small
\caption{Future events in the Universe of LVDP$t$ model.}
\vspace{0.1cm}
\begin{tabular}{ll}
\hline\hline Time & \quad \quad \quad Event\\ \hline
$t_{\rm BR}-23.1$ Gyr & \quad \quad \quad Today
\\
$t_{\rm BR}-0.53$ Gyr & \quad \quad \quad Erase Galaxy Clusters
\\
$t_{\rm BR}-31.53$ Myr & \quad \quad \quad Destroy Milky Way
\\
$t_{\rm BR}-1.58$ Months & \quad \quad \quad Unbind Solar System
\\
$t_{\rm BR}-4.30$ Days & \quad \quad \quad Strip Moon
\\
$t_{\rm BR}-13.32$ Minutes & \quad \quad \quad Earth Explodes
\\
$t_{\rm BR}-5\times 10^{-20}$ s & \quad \quad \quad Dissociate Atoms
\\
$t_{\rm BR}=36.60$ Gyr & \quad \quad \quad Big Rip
\\
\hline\hline
\end{tabular}
\label{tab:bigripresults}
\end{table}

From Table \ref{tab:bigripresults}, one may see that we are 23.1 Gyr away from the Big Rip moment. The dominant dark energy in phantom region ($w<-1$) will start erasing the galaxy clusters around 0.53 Gyr before the end of the Universe. Milky Way will be destroyed around 31.53 Myr while solar system will be disturbed around 1.58 months before the expiry of the Universe. Moon will be stripped apart around 4.30 days and destruction of earth will take place just before 13.32 minutes of the tragic end of the Universe. Finally the dissociation of atoms would happen in a fraction of the last second of the life of the Universe and thereby leading to the end of the Universe in Big Rip.

Here, we have discussed the fate of the Universe considering LVDP$t$ model. However, we should note that LVDP$t$ model is not obtained as a cosmological model based on a strong theoretical background, but as a model adopted from the first two terms of the Taylor series expansion of the unknown DP. Hence, the predicted fate of the Universe in the LVDP$t$ model should be considered just as an extrapolation to the future of the Universe in this particular model. For instance, it would be interesting to further investigate in a new study whether a Big Rip future will be favored or not when the LVDP$t$ model is extended by considering the first three terms in the Taylor series expansion of DP at the expense of low number number of free parameters to be constrained by the observations. However, on the other hand, we ultimately cannot be sure about the fate of the Universe as long as we do not experience the future. The most reliable but still not the ultimate prediction on the future of the Universe can be the one predicted through a successfully achieved theory of everything (such as string theory) that can pass all the experiments in the past of the Universe successfully.  However, we humankind cannot help asking the fate of the Universe and therefore, temptations would continue to discuss the fate of the Universe through the theoretical models that fit the observational data well. 

\begin{center}
\textbf{Acknowledgments}
\end{center}
{\small We thank to Bharat Ratra, Diego Pavon and S.D. Odintsov for fruitful comments on the paper.  \"{O}.A. and T.D. appreciate the support given by the Turkish Academy of Sciences (T{\"{U}}BA) while this research was carried out. \"{O}.A. acknowledges the postdoctoral research scholarship he is receiving from The Scientific and Technological Research Council of Turkey (T\"{U}B{\.I}TAK-B{\.I}DEB 2218). S.K. acknowledges the financial support from the Department of Science and Technology (DST), India under project No. SR/FTP /PS-102/2011. L. Xu's work is supported in part by NSFC under the Grant No. 11275035 and "the Fundamental Research Funds for the Central Universities" under the Grant No. DUT13LK01.}

\bigskip

\appendix
\section{Constraint methodology using observational Hubble ($H(z)$) data}
\label{sec:axHz}

The differential ages of the 
galaxies are used for the observational Hubble data via the relation  \cite{ref:JL2002,ref:JVS2003,ref:ZhangTJOHD} 
\begin{equation}
H(z)=-\frac{1}{1+z}\frac{dz}{dt}.
\end{equation} 
We have used 25 OHD data points shown in Table \ref{table2}, recently compiled by Zhang et al.\cite{ref:Zhang2012}. 

The mean values of the model parameters are determined by
minimizing
\begin{equation}
\chi_{OHD}^2(p_s)=\sum_{i=1}^{25} \frac{[H_{th}(p_s;z_i)-H_{
obs}(z_i)]^2}{\sigma_{H(z_i)}^2},\label{eq:chi2H}
\end{equation}
where $p_s$ denotes the parameters of the model, $H_{th}$
is the theoretical (model based) value for the Hubble parameter, $H_{obs}$ is the
observed one, $\sigma_{H(z_i)}$ is the standard error in the observed value, and the summation runs over the $25$ observational Hubble
data points at redshifts $z_i$.

\begin{center}
\begin{table}[h]\small\centering\footnotesize\caption{$H(z)({\rm km}\, {\rm s}^{-1}\,{\rm Mpc}^{-1})$ measurements and their $1\sigma$ errors.}\label{table2}
\begin{tabular}{llll}
\hline\hline\\
$z$ & $H(z)$  & $\sigma_{H(z)}$ & Reference \\\hline\hline
0.090  & 69    & 12    & Simon et al.\cite{ref:SVJ2005} \\
    0.170 & 83    & 8     & Simon et al.\cite{ref:SVJ2005} \\
    0.270 & 77    & 14    & Simon et al.\cite{ref:SVJ2005} \\
    0.400  & 95    & 17    & Simon et al.\cite{ref:SVJ2005} \\
    0.900 & 117   & 23    & Simon et al.\cite{ref:SVJ2005} \\
    1.300 & 168   & 17    & Simon et al.\cite{ref:SVJ2005} \\
    1.430 & 177   & 18    & Simon et al.\cite{ref:SVJ2005} \\
    1.530 & 140   & 14    & Simon et al.\cite{ref:SVJ2005} \\
    1.750 & 202   & 40    & Simon et al.\cite{ref:SVJ2005} \\
    0.24  & 79.69  &3.32  & Gaztanaga et al.\cite{Gaztanaga09}\\
    0.43  &86.45  &3.27  & Gaztanaga et al.\cite{Gaztanaga09}\\
    0.480 & 97    & 62    & Stern et al.\cite{stern10} \\
    0.880 & 90    & 40    & Stern et al.\cite{stern10} \\
    0.179 & 75    & 4     & Moresco et al.\cite{moresco12} \\
    0.199 & 75    & 5     & Moresco et al.\cite{moresco12} \\
    0.352 & 83    & 14    & Moresco et al.\cite{moresco12} \\
    0.593 & 104   & 13    & Moresco et al.\cite{moresco12} \\
    0.680 & 92    & 8     & Moresco et al.\cite{moresco12}\\
    0.781 & 105   & 12    & Moresco et al.\cite{moresco12} \\
    0.875 & 125   & 17    & Moresco et al.\cite{moresco12} \\
    1.037 & 154   & 20    & Moresco et al.\cite{moresco12} \\
0.07  & 69.0  & 19.6 & Zhang et al.\cite{ref:Zhang2012} \\
0.12  & 68.6  & 26.2 & Zhang et al.\cite{ref:Zhang2012}   \\
0.20  & 72.9  & 29.6 & Zhang et al.\cite{ref:Zhang2012} \\
0.28  & 88.8  & 36.6 & Zhang et al.\cite{ref:Zhang2012}  \\
\hline\hline\\
\end{tabular}
\end{table}
\end{center}

\section{Constraint methodology using observational SN Ia data}
\label{sec:axSN}

The observational SN Ia data have played an important role to discover the kinematic state of our Universe \cite{Riess98,Perlmutter99}.  Recently, Suzuki et al.\cite{ref:Suzuki2012} used the framework devised by Kowalski et al.\cite{ref:Kowalski2008} and  created a compilation of 580 supernovae, known as Union2.1 compilation. We use this Union2.1 SN Ia data compilation to constrain the model parameters as follows.  

The distance modulus $\mu(z)$ is defined as
\begin{equation}
\mu_{th}(z)=5\log_{10}[\bar{d}_{L}(z)]+\mu_{0},
\end{equation}
where $\mu_0\equiv42.38-5\log_{10}h$ with $H_0=100 h~{\rm km ~s}^{-1} {\rm Mpc}^{-1}$. Further, $\bar{d}_L(z)=H_0
d_L(z)/c$ is the Hubble-free luminosity, where 
\begin{eqnarray}
d_L(z)&=&\frac{c(1+z)}{H_0\sqrt{|\Omega_{k}|}}{\rm
S}\left[\sqrt{|\Omega_{k}|}\int^z_0\frac{dz'}{E(z')}\right]
\end{eqnarray}
with $E^2(z)=H^2(z)/H^2_0$, and $S(x)=\sinh x$ for $k=-1$,  $S(x)= x$ for $k=0$ and $S(x)=\sin x$ for $k=1$. The observed distance
modulus $\mu_{obs}(z_i)$ of SN Ia at redshift $z_i$ is given by
\begin{equation}
\mu_{obs}(z_i) = m_{obs}(z_i)-M,
\end{equation}
where $M$ is the absolute magnitude.

The mean values of the model parameters $p_s$ using the SN Ia data set
are determined by minimizing the quantity

\begin{eqnarray}
\chi^2(p_s,M^{\prime})&\equiv & \sum_{i=1}^{580}\frac{\left\{
\mu_{obs}(z_i)-\mu_{th}(p_s,z_i)\right\}^2} {\sigma_i^2}  \\\label{eq:chi2}
&=&\sum_{i=1}^{580}\frac{\left\{ 5 \log_{10}[\bar{d}_L(p_s,z_i)] -
m_{obs}(z_i) + M^{\prime} \right\}^2} {\sigma_i^2}, \nonumber
\end{eqnarray}
where $M^{\prime}\equiv\mu_0+M$ being a nuisance parameter  can be marginalized over
analytically \cite{ref:SNchi2} as

\begin{equation}
\bar{\chi}^2(p_s) = -2 \ln \int_{-\infty}^{+\infty}\exp \left[
-\frac{1}{2} \chi^2(p_s,M^{\prime}) \right] dM^{\prime},\nonumber
\label{eq:chi2marg}
\end{equation}
resulting to
\begin{equation}
\bar{\chi}^2 =  A - \frac{B^2}{C} + \ln \left( \frac{C}{2\pi}\right)
, \label{eq:chi2mar}
\end{equation}
with

\begin{eqnarray*}
A&=&\sum_{i,j}^{SN}\left\{5\log_{10}
[\bar{d}_L(p_s,z_i)]-m_{obs}(z_i)\right\}\cdot \\
& &{\rm
Cov}^{-1}_{ij}\cdot \left\{5\log_{10}
[\bar{d}_L(p_s,z_j)]-m_{obs}(z_j)\right\},\nonumber\\
B&=&\sum_{i,j}^{SN} {\rm Cov}^{-1}_{ij}\cdot \left\{5\log_{10}
[\bar{d}_L(p_s,z_j)]-m_{obs}(z_j)\right\},\nonumber \\
C&=&\sum_i^{SN} {\rm Cov}^{-1}_{ii}.\nonumber\label{eq:SNsyserror}
\end{eqnarray*}

In the above, ${\rm Cov}^{-1}_{ij}$ is the inverse of covariance matrix with or without systematic errors \cite{ref:Suzuki2012}. Relation
(\ref{eq:chi2}) has a minimum at 
$M^{\prime}=B/C$, which depends of the values of $h$
and $M$. Finally, the expression
\begin{equation}
\chi^2_{SN}(p_s)=A-\frac{B^2}{C},\label{eq:chi2SN}
\end{equation}
which matches (\ref{eq:chi2mar}) up to a constant, is often
used in the likelihood analysis \cite{ref:SNchi2}.
Thus, in this case the results would remain unaffected by a flat
$M^{\prime}$ distribution. It may be noted that the results will be
different with or without the systematic errors. In this study, all
results are derived considering the systematic errors.

\end{document}